\documentclass{aa}
\usepackage{graphicx}
\usepackage{amsmath, amssymb}
\usepackage{bm}
\usepackage[varg]{txfonts}
\usepackage{textcomp}
\usepackage{multirow}
\usepackage{natbib}
\usepackage{hyperref}

\bibpunct{(}{)}{;}{a}{}{,}

\begin{document}

\newcommand{\micron}{\textmu m}		% micron

\title{Mid-infrared observations of the circumstellar disks around PDS\,66 and CRBR\,2422.8-3423}

\author{C.~Gr\"afe\thanks{e-mail: cgraefe@astrophysik.uni-kiel.de}
	\and S.~Wolf
       }

\institute{
	    University of Kiel, Institute for Theoretical Physics and Astrophysics, Leibnizstrasse 15, 24118 Kiel, Germany 	\label{inst1}
	  }

\date{Received 22 January 2013 / Accepted 6 March 2013}

\abstract
  {}
  {We present mid-infrared observations and photometry of the circumstellar disks around \object{PDS\,66} and \object{CRBR\,2422.8-3423}, obtained with VISIR/VLT in the N band and for the latter also in the Q band. Our aim is to resolve the inner regions of these protoplanetary disks, which carry potential signatures of intermediate or later stages of disk evolution and ongoing planet formation.}
  {We determined the radial brightness profiles of our target objects and the corresponding PSF reference that were observed before and after our target objects. Background standard deviations, the standard errors, and the seeing variations during the observations were considered. Adopting a simple radiative transfer model based on parameters taken from previous studies, we derived constraints on the inner-disk hole radius of the dust disk.}
  {Neither of the circumstellar disks around our science targets are spatially resolved in our observations. However, we are able to constrain the inner-disk hole radius to $<\!15.0^{+0.5}_{-0.5}\,$AU and $<\!10.5^{+0.5}_{-1.0}\,$AU for PDS\,66 and CRBR\,2422.8-3423, respectively. The photometry we performed yields N-band flux densities of $599\pm8$\,mJy for PDS\,66 and $130\pm14$\,mJy for CRBR\,2422.8-3423, as well as a Q-band flux density of $858\pm109$\,mJy for CRBR\,2422.8-3423.}
  {}


\authorrunning{C.~Gr\"afe et al.}

\maketitle

\section{Introduction}
\label{section:introduction}

As a consequence of the conservation of angular momentum during star formation, young stars are generally surrounded by dust and gas-rich circumstellar disks. These disks are known to dissipate over time \citep[$\sim\!\!10^6$-$10^7\,$yr;][]{Haisch01} by means of stellar winds or photoevaporation by either the radiation of the central star or some external radiation \citep[e.g.,][]{Hollenbach00, Clarke01, Alexander07}, accretion onto the central star \citep{Hartmann98}, grain growth and fragmentation \citep[e.g.,][]{Dullemond05, Dominik07, Natta07, Birnstiel11, Sauter11, Garaud12, Ubach12}, and the formation of planets \citep[e.g.,][]{Pollack96, Boss02, Armitage07, Armitage10, Fortier12}. Knowledge of the structure of inner regions of disks and how they dissipate is crucial for our understanding of planet formation. Some protoplanetary disks show inner-disk clearings, such as gaps or holes, as an implication of the ongoing evolution of these disks. Their presence may indicate that planets have already formed in the disk and cleared their orbits. However, photoevaporation is also capable of producing inner-disk clearings. The study of such disks can therefore constrain models of planet formation and allows the roles of the different disk-clearing mechanisms to be evaluated. 
\par
To date, spectrophotometric methods are most widely used to identify inner-disk clearings \citep{Forrest04, Robitaille07, DAlessio09, Espaillat08, Espaillat10, Espaillat12}, but they usually do not provide an unequivocal solution. Conclusions obtained using SED modeling are much more model-dependent and degenerate than using resolved images. Therefore, imaging of protoplanetary disks provides an independent check of SED modeling and a far less ambiguous method for studying the disk structure \citep[e.g.,][]{Eisner09, Hughes09, Sauter09, Madlener12, Graefe12}. Various efforts to resolve the inner emission holes through direct imaging observations have been undertaken \citep[e.g. in the (sub)millimeter wavelength range by][]{Hughes07, Hughes09, Brown08, Brown09, Andrews11}. However, owing to their limited sensitivity, submillimeter and millimeter observations have so far only provided information about the large reservoirs of cold dust in the outer regions, and they hardly probe the inner hot regions since their thermal emission is very faint at (sub)mm wavelengths. Thus, high-resolution mid-infrared imaging of protoplanetary disks is essential for obtaining information about the warm dust in the innermost regions of the disk, for detecting inner-disk clearings as signatures of the disk dissipation mechanisms, and investigating the structure close to the star, hence the regions where planets potentially form \citep[e.g.,][]{Wolf07}. Efforts to spatially resolve circumstellar disks in the mid-infrared have been accomplished by, e.g., \citet{Pantin05}, \citet{Chen08}, \citet{Churcher11}, \citet{Graefe11}, and \citet{Honda12}.
\par
In this work we present high-resolution mid-infrared observations obtained with VISIR/VLT of the two well-studied circumstellar disks around PDS\,66 and CRBR\,2422.8-3423, reaching the limit of the angular resolution of today's single-dish, ground-based telescopes. An even higher resolution in this wavelength range can only be achieved by means of interferometric observations, e.g., with MIDI/VLTI \citep[e.g.,][]{Schegerer09}, which are still limited to the mid-infrared brightest sources. Our target objects are characterized in \S\,\ref{section:target_object}, while the observations and data reduction are found in \S\,\ref{section:obs_datared}. In \S\,\ref{section:data_analysis} we describe the analysis of the data and the applied model in detail and finally present and discuss our results in \S\,\ref{section:results_discussion}.
\par

\section{Target objects}
\label{section:target_object}

  \subsection{PDS\,66}
  \label{subsection:pds66}

\begin{figure}[!htb]
 \centering
  \includegraphics[width=0.50\textwidth]{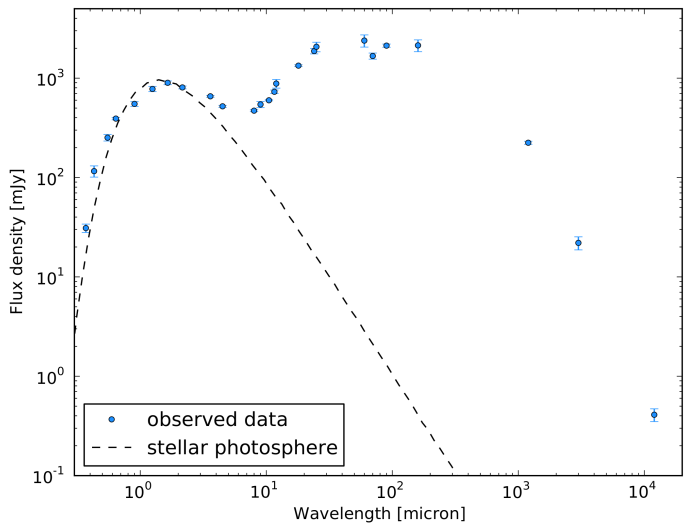}
\caption{Spectral energy distribution of PDS\,66. A star with $T_\star = 5035\,$K is shown as the dashed line.}
\label{Fig:pds66_sed}
\end{figure}

PDS\,66, also known as MP\,Mus and Hen\,3-892, is a K1\,IVe classical T\,Tauri star (cTTS) and a member of Lower Centaurus Crux \citep{Gregorio92, Mamajek02}. It exhibits excess infrared emission at $\lambda > 3\,$\micron~that has been interpreted as arising from a primordial circumstellar disk \citep[e.g.,][]{Silverstone06, Hillenbrand08}. A distance $d$ of only $86\,$pc makes it to one of the nearest known protoplanetary disks \citep{Mamajek02}. Despite its old age of $\sim\!13\,$Myr, which is substantially beyond the median age for cTTS \citep[e.g.,][]{Pascucci07, Cortes09}, the disk of PDS\,66 is in a stage between a primordial and a transitional disk, evolving toward the transition disk population, not showing indications of any large cleared inner regions in its spectral energy distribution (SED) yet \citep[e.g.,][]{Preibisch08, Cortes09}. It is still actively accreting with a mass accretion rate of $5\times10^{-9}{\rm M_\odot\,yr^{-1}}$ resulting from modeling the H$\alpha$ line profile \citep{Pascucci07}. Figure~\ref{Fig:pds66_sed} shows the SED of PDS\,66. In Table~\ref{Tab:pds66_fluxes} all available flux measurements are compiled.
\par

\begin{table}[!htb]
\caption{Flux densities of PDS\,66 as shown in Fig.~\ref{Fig:pds66_sed}}
\label{Tab:pds66_fluxes}
\centering
\begin{tabular}{l c c c}
\hline\hline
$\lambda$ [\micron]	&	Flux density [mJy]	&	Instrument	&	Reference	\\
\hline
0.37			&	$31\pm3$		&			&	(1)		\\
0.43			&	$116\pm15$		&	Hipparcos	&	(2)		\\
0.55			&	$252\pm19$		&	Hipparcos	&	(2)		\\
0.64			&	$392\pm11$		&			&	(1)		\\
0.9			&	$551\pm30$		&	La Silla/ESO 1m	&	(3)		\\
1.25			&	$779\pm46$		&	2MASS facility	&	(4)		\\
1.65			&	$899\pm39$		&	2MASS facility	&	(4)		\\
2.16			&	$807\pm27$		&	2MASS facility	&	(4)		\\
3.6			&	$657\pm14$		&	SST/IRAC	&	(5)		\\
4.5			&	$521\pm12$		&	SST/IRAC	&	(5)		\\
8.0			&	$471\pm10$		&	SST/IRAC	&	(5)		\\
9.0			&	$544\pm39$		&	AKARI		&	(6)		\\
10.5			&	$599\pm8$		&	VLT/VISIR	&	(7)		\\
11.6			&	$730\pm30$		&	La Silla/TIMMI2	&	(8)		\\
12			&	$882\pm88$		&	IRAS		&	(9)		\\
18			&	$1337\pm27$		&	AKARI		&	(6)		\\
24			&	$1874\pm120$		&	SST/MIPS	&	(5)		\\
25			&	$2070\pm228$		&	IRAS		&	(9)		\\
60			&	$2390\pm335$		&	IRAS		&	(9)		\\
70			&	$1672\pm118$		&	SST/MIPS	&	(5)		\\
90			&	$2126\pm79$		&	AKARI		&	(6)		\\
160			&	$2138\pm290$		&	SST/MIPS	&	(5)		\\
1200			&	$224\pm8$		&	SEST/SIMBA	&	(8)		\\
3000			&	$22\pm3.3$		&	ATCA		&	(1)		\\
12000			&	$0.41\pm0.06$		&	ATCA		&	(1)		\\
\hline
\end{tabular}
\tablebib{(1) \citet{Cortes09}; (2) Tycho-2 Catalog; (3) DENIS database; (4) 2MASS All-Sky Catalog of Point Sources; (5) \citet{Hillenbrand08}; (6) AKARI Point Source Catalogs; (7) This work; (8) \citet{Schuetz05a}; (9) IRAS Catalog of Point Sources}
\tablefoot{SST: Spitzer Space Telescope; IRAC: Infrared Array Camera; VLT: Very Large Telescope; VISIR: VLT spectrometer and imager for the mid-infrared; TIMMI2: Thermal Infrared MultiMode Instrument; IRAS: Infrared Astronomical Satellite; MIPS: Multiband Imaging Photometer for Spitzer; SEST: Swedish ESO Submillimeter Telescope; SIMBA: SEST IMaging Bolometer Array; ATCA: Australia Telescope Compact Array\\
In addition to the flux densities listed above there is also a TIMMI2 spectrum in the range [8, 13]\,\micron~\citep{Schuetz05a}, a SST/Infrared Spectrograph low-resolution spectrum in the range [5, 35]\,\micron~\citep{Silverstone06, Bouwman08}, and a spectrum in the range [0.3, 1]\,\micron~obtained with the 4m Cerro Tololo Inter-American Observatory \citep{Pascucci07} available in the literature.}
\end{table}

  \subsection{CRBR\,2422.8-3423}
  \label{subsection:crbr2422}

\begin{figure}[!htb]
 \centering
  \includegraphics[width=0.50\textwidth]{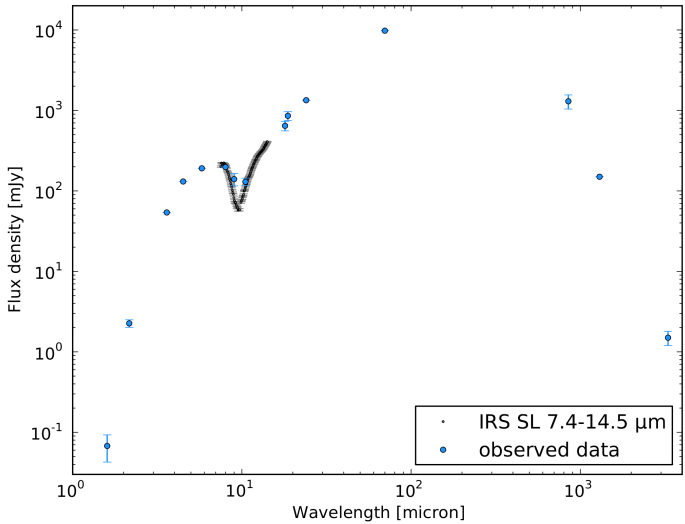}
\caption{Spectral energy distribution of CRBR\,2422.8-3423 including a SST/IRS spectrum in the range $7.4-14.5$\,\micron~\citep{Pontoppidan05}.}
\label{Fig:crbr_sed}
\end{figure}

CRBR\,2422.8-3423 is a Class I source located in the $\rho$\,Ophiuchus dark cloud at a distance of $120\,$pc \citep{Comeron93, Motte98, Brandner00}. Its near-infrared appearance is dominated by a bipolar reflection nebula intersected by a central dust lane, similar to the well-studied edge-on disk of the Butterfly Star in Taurus \citep{Brandner00, Pontoppidan05, Graefe12}. The southeastern lobe of the reflection nebula is 11 times fainter than the northwestern lobe in the K band \citep{Brandner00}, showing that the disk has an inclination deviating $\sim\!20^\circ$ from edge-on orientation \citep{Thi02}. No signatures of accretion or outflow are detected. \citet{Pontoppidan05} detected ices in the disk using spectroscopy obtained with the Spitzer Space Telescope. Due to the lower than edge-on disk inclination of CRBR\,2422.8-3423, the upper parts of the disk become optically thinner in the mid-infrared along the line of sight and allow one the detection of continuum emission from the warm, inner disk. Figure~\ref{Fig:crbr_sed} shows the SED of CRBR\,2422.8-3423 with its strong infrared excess resulting from the circumstellar disk. In Table~\ref{Tab:crbr_fluxes} the available flux measurements are compiled.
\par

\begin{table}[!htb]
\caption{Flux densities of CRBR\,2422.8-34236 as shown in Fig.~\ref{Fig:crbr_sed}}
\label{Tab:crbr_fluxes}
\centering
\begin{tabular}{l c c c}
\hline\hline
$\lambda$ [\micron]	&	Flux density [mJy]	&	Instrument	&	Reference	\\
\hline
1.6			&	$0.07\pm0.03$		&	VLT/ISAAC	&	(1)		\\
2.16			&	$2.26\pm0.24$		&	2MASS facility	&	(2)		\\
3.6			&	$54$			&	SST/IRAC	&	(3)		\\ 
4.5			&	$131$			&	SST/IRAC	&	(3)		\\
5.8			&	$191$			&	SST/IRAC	&	(3)		\\
8.0			&	$198$			&	SST/IRAC	&	(3)		\\
9.0			&	$140\pm25$		&	AKARI		&	(4)		\\
10.5			&	$130\pm14$		&	VLT/VISIR	&	(5)		\\
18.0			&	$645\pm88$		&	AKARI		&	(4)		\\
18.7			&	$858\pm109$		&	VLT/VISIR	&	(5)		\\
24			&	$1340$			&	SST/MIPS	&	(3)		\\
70			&	$9800$			&	SST/MIPS	&	(3)		\\
850			&	$1300\pm260$		&	JCMT/SCUBA	&	(3)		\\
1300			&	$150$			&	IRAM 30m	&	(6)		\\
3300			&	$1.5\pm0.3$		&	ATCA		&	(7)		\\
\hline
\end{tabular}
  \tablebib{(1) \citet{Brandner00}; (2) 2MASS All-Sky Catalog of Point Sources; (3) \citet{vanKempen09}; (4) AKARI Point Source Catalogs; (5) This work; (6) \citet{Motte98}; (7) \citet{Ricci10}}
\tablefoot{ISAAC: Infrared Spectrometer And Array Camera; JCMT: James Clerk Maxwell Telescope; SCUBA: Submillimetre Common-User Bolometer Array; IRAM: Institut de Radioastronomie Millim\'etrique\\
In addition to the flux densities listed above there is also a K-band spectrum \citep{Brandner00}, a M-band spectrum \citep{Thi02}, and a SST/IRS spectrum \citep{Pontoppidan05} available in the literature.}
\end{table}

\section{Observations and data reduction}
\label{section:obs_datared}

The mid-infrared observations of our two targets, PDS\,66 and CRBR\,2422.8-3423, were carried out on February 24 and May 21, 2011, respectively, using the Very Large Telescope (VLT) spectrometer and imager for the mid-infrared (VISIR) \citep{VISIR04} in the SIV filter and for CRBR\,2422.8-3423 also in the Q2 filter. The properties of the observations are listed in Table~\ref{Tab:obs_prop}. To achieve the highest possible angular resolution, the small field objective was used (0.075\arcsec/pixel). A standard chop/nod-procedure to remove sky and telescope emission with a chop of 8$\arcsec$ in the north-south direction and a perpendicular 8$\arcsec$ nod as well as a jitter width of 1\arcsec~were applied for the observations. Calibration observations of standard stars within a few degrees of the science object were taken immediately before and after the science observations. The standard stars were chosen from the mid-infrared spectrophotometric standard star catalog of the VLT based on \citet{Cohen99}. All observations were executed in service mode.
\par

\begin{table*}[!htb]
\caption{Observation properties}
\label{Tab:obs_prop}
\centering
\begin{tabular}{l c c c c c c c}
\hline\hline
Target			& Type	& RA (J2000)	& Dec (J2000)	& Observing Date & Exposure	 & Filter	& $\lambda_{\rm c}$	\\
			&	& [hh mm ss]	& [dd mm ss]	& [2011]	 & Time [s]	 & 		& [\micron]		\\
\hline
HD\,91056		& STD	& 10 28 52.30	& -64 10 26.1	& Feb 24	 & $2\times62.4$ & SIV		& $10.49$		\\
PDS\,66			& SCI	& 13 22 07.55	& -69 38 12.2	& Feb 24	 & $1794.0$	 & SIV		& $10.49$		\\

\multirow{2}{*}{HD\,139663}		& \multirow{2}{*}{STD}	& \multirow{2}{*}{15 40 16.80}	& \multirow{2}{*}{-23 49 19.5}	& \multirow{2}{*}{May 21}	 									 & $2\times62.4$ & SIV		& $10.49$		\\
			&	&		&		&		 & $2\times95.0$ & Q2		& $18.72$		\\

\multirow{2}{*}{CRBR\,2422.8-3423} 	& \multirow{2}{*}{SCI}	& \multirow{2}{*}{16 27 24.61}	& \multirow{2}{*}{-24 41 03.3}	& \multirow{2}{*}{May 21}	 									 & $717.6$	 & SIV		& $10.49$		\\
			&	&		&		&		 & $342.0$	 & Q2		& $18.72$		\\
\hline
\end{tabular}
\tablefoot{SCI: science object, STD: standard star, $\lambda_{\rm c}$: central filter wavelength}
\end{table*}

Raw data were reduced using the VISIR data reduction pipeline provided by ESO. A full-resolution background map was created for every observed image using the Source Extractor software \citep{Bertin96} and subtracted from the image to remove the background. This approach provides a more reliable estimate of the background than the assumption of a constant one. Owing to the small chopping and nodding throw in the observations, the observed source is always present in the field of view of the instrument such that the source is detected four times in the reduced image. To increase the signal-to-noise ratio ($S/N$) by a factor of two, these four detections of the source were averaged. The resulting reduced and averaged images of our two targets as well as the corresponding standard star images were used for the data analysis (\S\,\ref{section:data_analysis}).
\par

\section{Data analysis}
\label{section:data_analysis}

The data analysis presented here is similar to previous work \citep{Graefe11}. For both target objects the mid-infrared emission is dominated by the circumstellar disk.

  \subsection{Radial brightness profiles}
  \label{subsection:profiles}

If the circumstellar disk surrounding the central star is spatially resolved, we expect the observed brightness distribution of this object to be significantly broadened compared to the corresponding point spread function (PSF). The observed standard stars are considered to be point sources and therefore used as PSF references in the data analysis. An appropriate comparison between science object and corresponding standard star observation clarifies if the circumstellar disk is spatially resolved.
\par
We characterize the brightness distribution of an observed object by its peak-normalized azimuthally averaged radial brightness profile. The averaging is performed with respect to the center of each object which is determined by fitting a two-dimensional Gaussian profile. Assuming azimuthal symmetry, this averaging method yields a much better brightness profile than a simple cut through the object and also considerably improves the $S/N$ by a factor of $\sqrt{n}$, where $n$ is the number of pixels in every annulus. Azimuthally symmetric disk images can be expected if the disks are seen in face-on orientation or if they are unresolved but the PSF is approximately circular. The effect of the disk inclination on our results is discussed in \S\,\ref{section:results_discussion}.
\par
We compile a radial brightness profile for each science object and standard star. For this purpose, the two standard star observations of each science object observation are averaged. Subtracting the standard star profile from the corresponding science object profile results in a residual flux profile produced by the extended star-disk system. Significant residual flux implies that the circumstellar disk is spatially resolved. 
\par

  \subsection{Detection threshold}
  \label{subsection:det_thrshld}

To investigate whether the derived residual flux is significant, we define a detection threshold such that any residual flux above this threshold is caused by an extension of the star-disk system, i.e., a spatially resolved disk surrounding the star.
\par
Our detection threshold takes into account the standard deviation of the background flux (background noise, $\sigma_{\rm bg}$) of the science and standard star image, respectively, each standard error ($\sigma_{\rm SE}$) resulting from the averaging of the brightness profiles of both images as well as the seeing variations during the observation. Owing to the azimuthal averaging, the background noise of the brightness profiles depends on the radius ($r$) and declines with increasing radius. We impose the constraint that the residual flux must be higher than $3\sigma_{\rm bg}$. The standard error is also a function of $r$ and is determined by $\sigma_{\rm SE} = \sigma/\sqrt{n}$ where $\sigma$ is the standard deviation of the averaged flux of the brightness profile. The seeing which varies during observations increases the full-width-at-half-maximum (FWHM) of the observed object. In the data analysis, such variations can cause a residual flux profile similar to the case of an extended star-disk system, thus mimic a spatially resolved disk. Therefore, it is crucial to determine these variations during the observation. To estimate the seeing variations we observed the standard stars before and after the science object. Assuming that the seeing has a normal distribution during the observations, the variations can be calculated as the half of the absolute difference of the residual flux profiles of both standard star observations. This method yields a radial profile reflecting a conservative estimate of the seeing variations during the observations. In addition to the already mentioned constrains, the derived residual flux profiles are considered significant only if they are three times larger than the derived seeing variations.
\par
\begin{figure}[!htb]
 \centering
  \includegraphics[width=0.50\textwidth]{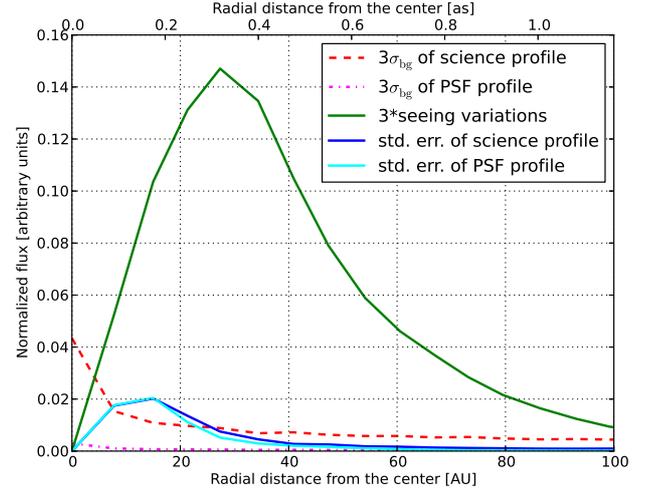}
\caption{Contribution of each of the components to the detection threshold for PDS\,66.}
\label{Fig:det_thrshld}
\end{figure}

Our detection threshold is now defined as the sum of $3\sigma_{\rm bg}$ and $\sigma_{\rm SE}$ of the science and standard star profile, respectively, and three times the seeing variations. The contribution of each of the components to this threshold is exemplarily represented for PDS\,66 in Fig.~\ref{Fig:det_thrshld}, emphasizing the necessity for a reliable determination of the seeing variations.
\par

  \subsection{Model}
  \label{subsection:model}

During the evolution of circumstellar disks, inner-disk clearings like holes are expected to form (see also \S\,\ref{section:introduction}). Such objects are often referred to as transitional disks and are considered to represent an evolved stage of the accretion disks. In the mid-infrared, the surface brightness distribution of the circumstellar disk is dominated by the inner disk rim, which is directly illuminated by the central star. It is our aim to give constraints on the inner-disk hole radius of the observed objects adopting a simple model based on parameters taken from previous studies.
\par
Our model consists of a parameterized disk that is described with a radially and vertically dependent density distribution based on the work of \citet{Shakura73}:
\begin{gather}
\rho_{\rm disk} = \rho_0 \left(\frac{r_0}{r_{\rm cyl}}\right)^{\alpha} {\rm exp}\left(-\frac{1}{2}\left[\frac{z}{h(r_{\rm cyl})}\right]^2\right).
\label{Eq:density}
\end{gather}
Here, z is the cylindrical coordinate with $z=0$ corresponding to the disk midplane and $r_{\rm cyl}$ is the radial distance from this $z$-axis. The parameter $\rho_0$ is determined by the total dust mass $m_{\rm dust}$ of the disk. The quantity $r_0$ is a reference radius and has a typical value of $100\,$AU whereas $h$ is the vertical scale height and a function of $r_{\rm cyl}$
\begin{gather}
h(r_{\rm cyl}) = h_0 \left(\frac{r_{\rm cyl}}{r_0}\right)^{\beta}.
\label{Eq:scale_height}
\end{gather}
The parameters $\alpha$ and $\beta$ describe the radial density profile and the disk flaring, respectively. The disk extends from an inner cylindrical radius $r_{\rm in}$ to an outer one $r_{\rm out}$. This ansatz has already been successfully applied to other circumstellar disks, such as, e.g., the Butterfly Star \citep{Graefe12}, HH\,30 \citep{Madlener12}, and CB\,26 \citep{Sauter09}. Integrating Eq.~\ref{Eq:density} along the $z$-axis yields the surface density
\begin{gather}
\Sigma\left(r_{\rm cyl}\right) = \Sigma_0\left(\frac{r_{\rm cyl}}{r_0}\right)^{-p} {\hspace{0.4cm}\rm with \hspace{0.4cm}} p = \alpha - \beta.
\label{Eq:surf_den}
\end{gather}
The density structure described in Eq.~\ref{Eq:density} is given by the gas in the disk. As the dust in the disk is assumed to be coupled to the gas, its density distribution is also described by this equation. The sizes of the dust grains in our model are distributed according to a power law of the form
\begin{gather}
{\rm d}n\left(a\right) \sim a^{-3.5}\,{\rm d}a {\hspace{0.4cm}\rm with \hspace{0.4cm}} a_{\rm min} < a < a_{\rm max}.
\label{Eq:grains}
\end{gather}
Here, $a$ is the dust grain radius and $n(a)$ the number of dust grains with a specific radius. The parameters used in our model are listed in Table~\ref{Tab:model}. Most of them are taken from previous studies of both objects. For the grain size distribution we use the commonly known MRN distribution by \citet{Mathis77} and we fix the disk scale height at 100\,AU at a typical value of 10\,AU \citep[e.g.,][]{Graefe12, Sauter09, Pinte08}. Using the relation $\alpha = 3(\beta - 0.5)$, which results from viscous accretion theory \citep{Shakura73}, and the given surface density exponent $p = \alpha - \beta$, the parameters $\alpha$ and $\beta$ amount to $2.25$ and $1.25$, respectively. The inner disk radius is used as variable parameter.
\par
To allow us a comparison with the observations, the simulated images, that are constructed using the radiative transfer code \texttt{MC3D} \citep{Wolf99, Wolf03MC3D}, are convolved with the corresponding standard star observations. Following the method described in \S\,\ref{subsection:profiles} a residual flux profile is determined for every synthetic image. If the disk is spatially resolved in the observation we can reproduce the residual flux profile with the described model by varying the inner disk radius and thus determine the inner-disk hole radius of the object. For disks that could not be spatially resolved our model allows us to estimate an upper limit to the inner-disk hole radius. 

\begin{table}[!htb]
\caption{Adopted model parameters.}
\label{Tab:model}
\centering
\begin{tabular}{l c c c c}
\hline\hline
\multirow{2}{*}{Parameter}			& \multicolumn{2}{c}{PDS\,66}	& \multicolumn{2}{c}{CRBR\,2422.8-3423} \\
				& Value		& Reference	& Value		& Reference   \\
\hline
$d$ [pc]			& 86		& (1)		& 120		& (4)	\\
$T_\star$ [K]			& 5035		& (1)		& 4500		& (3)	\\
$L_\star$ [$\rm{L_\odot}$]	& 1.1		& (1)		& 1.4		& (3)	\\
$m_{\rm dust}$ [$\rm{M_\odot}$]	& 0.00005	& (2)		& 0.000015	& (3)	\\
$a_{\rm min}$ [nm]		& 5		& (5)		& 5		& (5)	\\
$a_{\rm max}$ [nm]		& 250		& (5)		& 250		& (5)	\\
$r_{\rm out}$ [AU]		& 170		& (1)		& 90		& (3)	\\
$i$ [$^\circ$]			& 32		& (1)		& 70		& (3)	\\
$h_0$ [AU]			& 10		& (5)		& 10		& (5)	\\
$p$				& 1		& (1)		& 1		& (3)	\\
$\alpha$			& 2.25		& (5)		& 2.25		& (5)	\\
$\beta$				& 1.25		& (5)		& 1.25		& (5)	\\
\hline
\end{tabular}
\tablebib{(1) \citet{Cortes09}; (2) \citet{Carpenter05}; (3) \citet{Pontoppidan05}; (4) \citet{vanKempen09}; (5) see text}
\end{table}

\par

  \subsection{Photometry}
  \label{subsection:photometry}

The flux densities of the two observed objects are determined using the VISIR photometry package by \citet{Hoenig10}. For photometry, the science targets can be considered as point sources such that each science and calibrator beam is fitted by a two-dimensional Gaussian. The flux densities are then obtained by calculating conversion factors from the integrated intensity of the Gaussian fits of the calibrators and using these factors to evaluate the respective integrated Gaussian intensity of the science targets.
\par

\section{Results and discussion}
\label{section:results_discussion}

The resulting constraints on the inner-disk hole radii and the photometry of the observed circumstellar disks are compiled in Table~\ref{Tab:results}. In the following, we discuss the analysis of PDS\,66 and CRBR\,2422.8-3423 in detail.\par

\begin{table}[!htb]
\caption{Constraints on the inner-disk hole radii and photometry of the observed objects}
\label{Tab:results}
\centering
\begin{tabular}{l c c c}
\hline\hline
Target			& $R_{\rm in}$ [AU]	&	Flux density [mJy]	& $\lambda_{\rm c}$ [\micron]	\\
\hline
PDS\,66			& $<\!15.0^{+0.5}_{-0.5}$	&	$599\pm8$	& $10.49$		\\
\multirow{2}{*}{CRBR\,2422.8-3423}	& $<\!10.5^{+0.5}_{-1.0}$	&	$130\pm14$	& $10.49$	\\
			& --			&	$858\pm109$	& $18.72$			\\
\hline
\end{tabular}
\tablefoot{The uncertainties in the inner-disk hole radii are caused mainly by the uncertainty in the disk inclination (see \S\,\ref{section:results_discussion}).}
\end{table}

\begin{figure*}[!htb]
 \centering
  \includegraphics[width=0.49\textwidth]{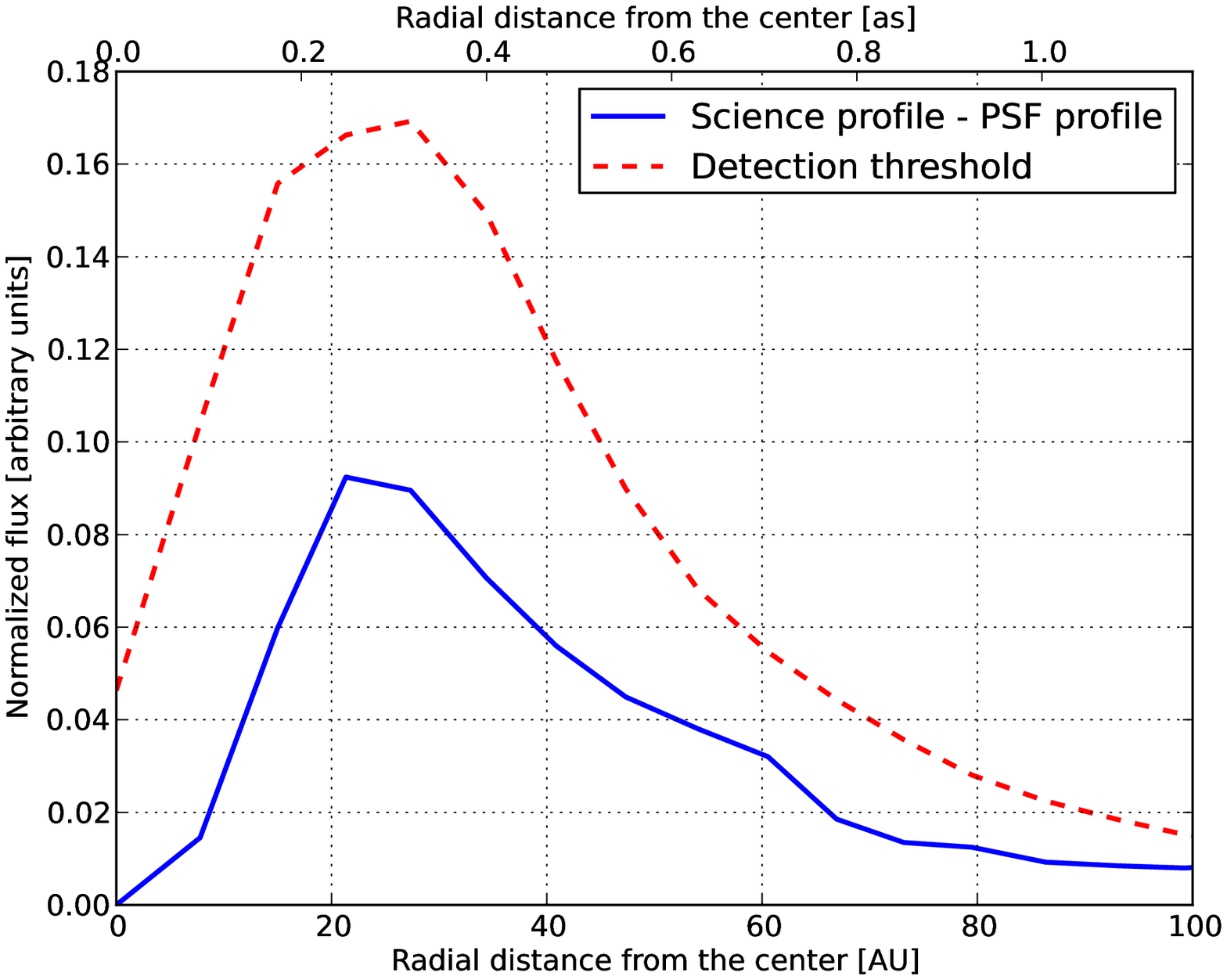}
  \hfill
  \includegraphics[width=0.49\textwidth]{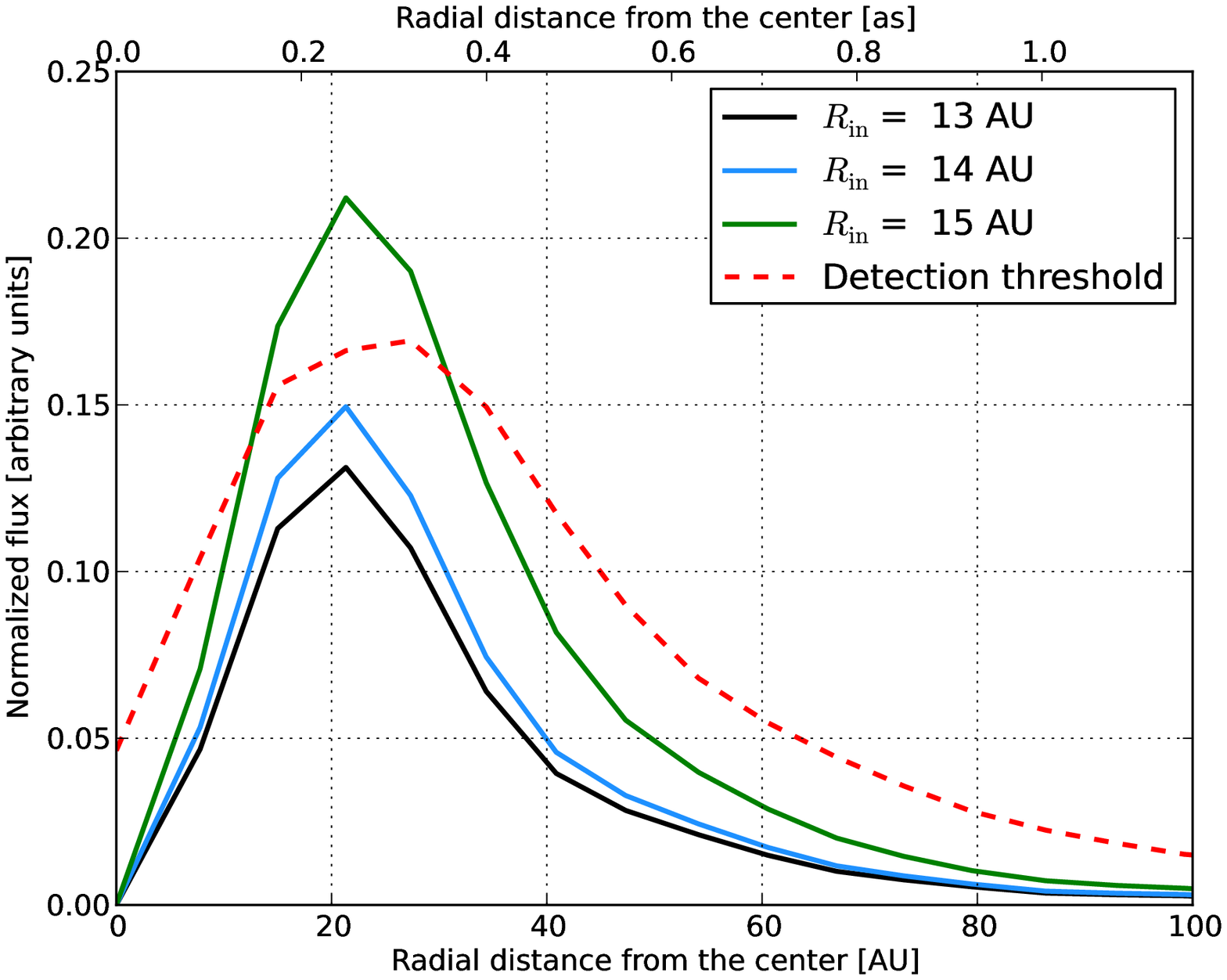}
\caption{\emph{Left}: Observed residual flux of {\bf PDS\,66} including the determined detection threshold. \emph{Right}: Modeled residual flux for several inner-disk hole radii including the detection threshold.}
\label{Fig:pds66_results}
\end{figure*}

\begin{figure*}[!htb]
 \centering
  \includegraphics[width=0.49\textwidth]{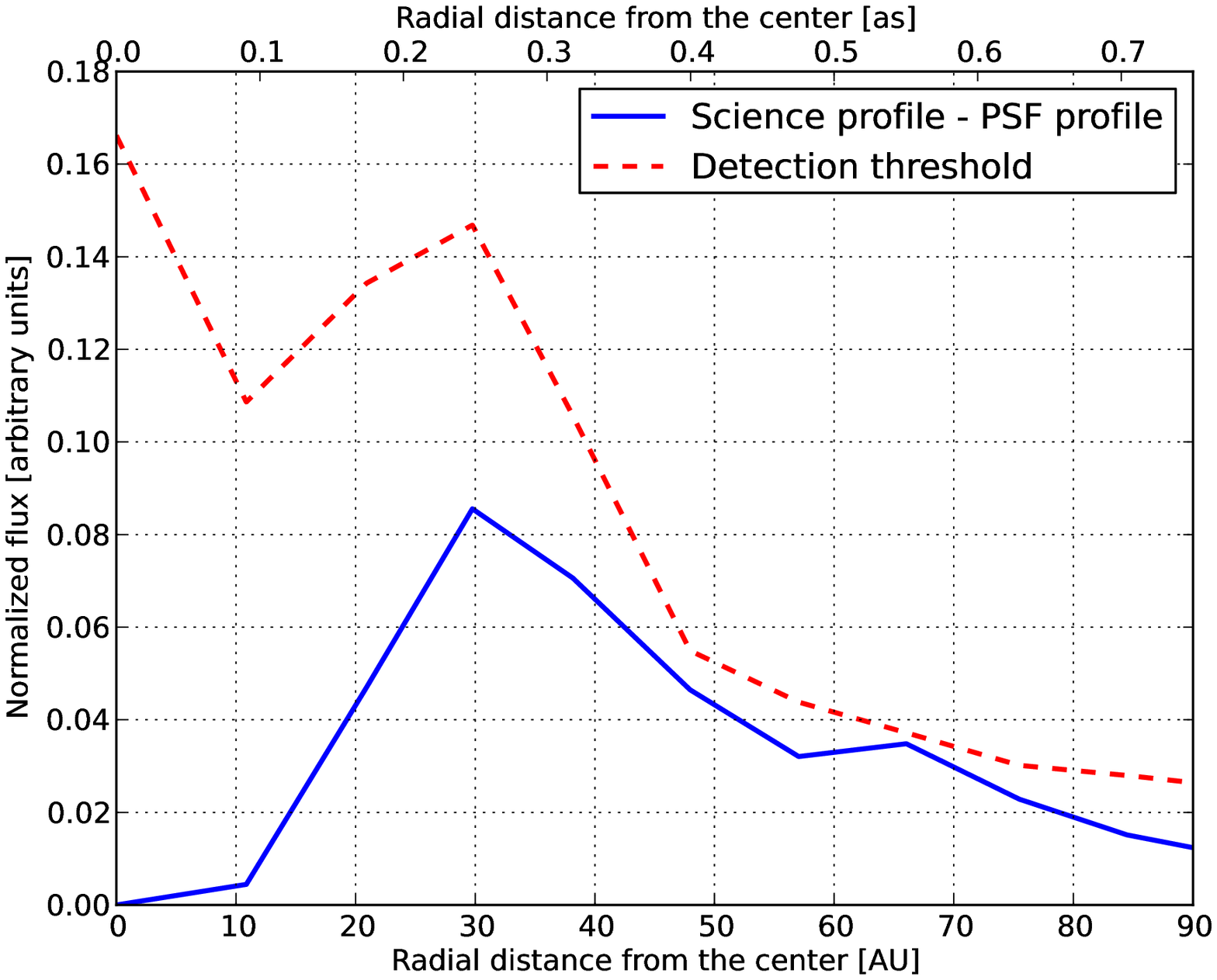}
  \hfill
  \includegraphics[width=0.49\textwidth]{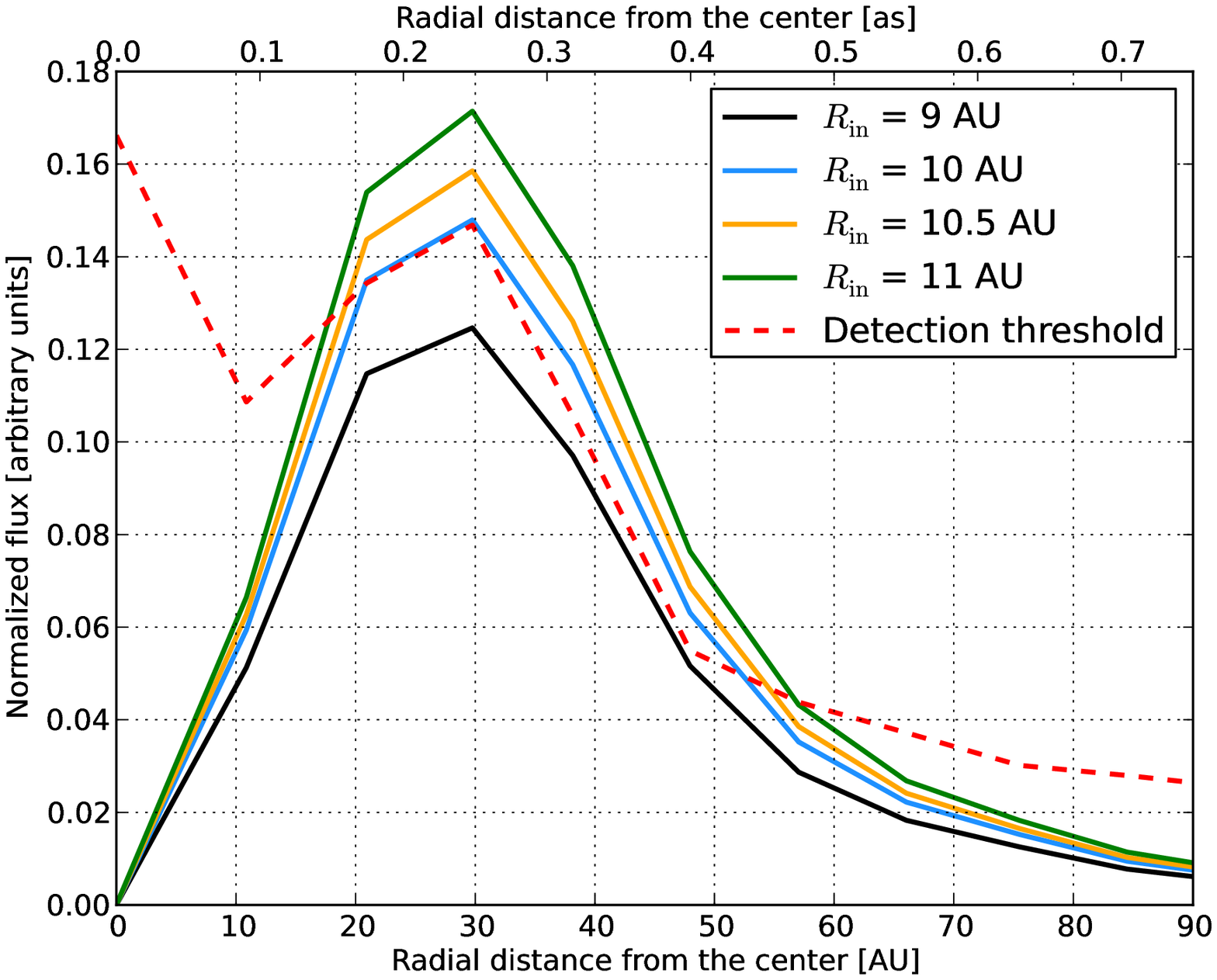}
\caption{\emph{Left}: Observed residual flux of {\bf CRBR\,2422.8-3423 N band} including the determined detection threshold. \emph{Right}: Modeled residual flux for several inner-disk hole radii including the detection threshold.}
\label{Fig:crbr_results}
\end{figure*}

For both objects, the difference in the object and standard star flux profiles is below the detection threshold, i.e., no significant extension of the star-disk system is detected (see Fig.~\ref{Fig:pds66_results}, left and Fig.~\ref{Fig:crbr_results}, left). Using our model, we can investigate at which inner-disk hole radius we would have detected a significant extension of the star-disk system, thus determine an upper limit to the inner-disk hole radius. We infer the inner radii to be smaller than $15.0^{+0.5}_{-0.5}\,$AU and $10.5^{+0.5}_{-1.0}\,$AU for PDS\,66 and CRBR\,2422.8-3423, respectively (see Fig.~\ref{Fig:pds66_results}, right and Fig.~\ref{Fig:crbr_results}, right). The uncertainties result from the uncertainty of the disk inclination which is $32\degr\pm5\degr$ for PDS\,66 \citep{Cortes09} and $70\degr\pm5\degr$ for CRBR\,2422.8-3423 \citep{Thi02}. The constraints on the inner-disk hole radii we give here are the first for these two objects and consistent with what is known about both objects from the literature. The quality of the Q-band observation of CRBR\,2422.8-3423 and also the corresponding standard star is too low to be used for our analysis. This is also reflected in the uncertainty of the flux measurement (Table~\ref{Tab:results}). Therefore, we do not give any constraints on the inner-disk hole radius of CRBR\,2422.8-3423 resulting from the Q-band observations.
\par
Owing to its old age, the disk around PDS\,66 is very likely to already have formed an inner-disk hole or at least inner regions that are depleted. Unfortunately, this could not be confirmed by our observations. Nevertheless, our analysis gives an upper limit on the inner-disk hole and future observations will hopefully resolve or further constrain it. In contrast to PDS\,66, CRBR\,2422.8-3423 is a Class I object, thus its disk is not expected to already have created large inner-disk clearings. This is also supported by our analysis providing a low upper limit for an inner-disk hole.
\par
As shown in Fig.~\ref{Fig:det_thrshld} the seeing variations are the major component of the detection threshold. A proper estimate of them is crucial to prevent the detection of a supposed extended star-disk system that is not caused by a circumstellar disk but by variations in the observing conditions during observation leading to a broadening of the brightness profile.
\par
The diffraction limit for VLT/VISIR in the N band is $\sim\!0.3\arcsec$. For PDS\,66 with a distance of $86\,$pc this means that it is in principle possible to spatially resolve structures with a radius as small as $13\,$AU. Our upper limit on the inner-disk hole radius of $15.0^{+0.5}_{-0.5}\,$AU is just above the diffraction limit. For CRBR\,2422.8-3423 with a distance of $120\,$pc, structures of $18\,$AU in radius can be spatially resolved at the diffraction limit. However, we determine an upper limit for the inner-disk hole radius of $10.5^{+0.5}_{-1.0}\,$AU which is below the diffraction limit and thus seems odd at first. This result can be easily explained by the fact that the location of the inner-disk radius influences the temperature distribution on larger scales. Therefore, based on the underlying disk model (density distribution, dust properties), the location of the inner rim can be constrained from the mid-infrared brightness distribution on larger scales. However, this statement only applies if the observed brightness distribution is not solely given by the inner rim (face-on orientation) but dominated by the re-emission of these outer regions. The latter is the case for the disk around CRBR\,2422.8-3423 which is seen at an inclination of $70\degr$, where the inner rim cannot be seen directly.
\par
As described in \S\,\ref{subsection:profiles} we characterize the brightness distribution of an observed object by its peak-normalized azimuthally averaged radial brightness profile. This method can be used without hesitation for stars surrounded by face-on orientated disks. However, this method can also be applied to inclined disks yielding a more conservative result for the broadening of the object with respect to the corresponding PSF. Since the model uses the same inclination and PSF as the observation, a comparison between both is feasible with this method.
\par
The photometry for our two observed objects yields, in the N band, a flux density of $599\pm8$\,mJy for PDS\,66 and $130\pm14$\,mJy for CRBR\,2422.8-3423 and in the Q2 band $858\pm109$\,mJy for CRBR\,2422.8-3423. As a complement to already known flux densities at other wavelengths, they will be useful for constraining the SEDs of these two objects and modeling their circumstellar disks.
\par

\begin{acknowledgements}
This work is supported by the DFG through the research group 759 ''The Formation of Planets: The Critical First Growth Phase'' (WO 857/4-2) and project WO 857/12-1. We are grateful to the anonymous referee for providing useful suggestions that improved this paper.
\end{acknowledgements}

\bibliographystyle{aa}
\bibliography{bibfile}

\end{document}